# About the wording of Fermat`s principle for light propagation in media with negative refraction index.

Veselago V.G.

*Moscow Institute of Physics and Technology, 141700, Dolgoprudhy, Russia*
*A.M.Prokhorov General Physics Institute of RAS*

In this work is elaborated the wording of Fermat`s principle for electromagnetic waves spreading in materials with negative index of refractions (MNR). It is shown that for MNR not applicable wording of Fermat`s principle through minimum (or extremum) of time of spreading the wave between two points. The correct wording for this case is a wording through extremum of total optical length for way between two points, assuning negative sign for the optical length if wave passing through MNR.

The papers under the group of scientists from University San-Diego [1,2] communicate about practical realization of composite material, unusual electrodynamic properties which can be well explained if assume that index of refraction of such material is negative. Hereunder, there were experimental confirmed main positions of work [3]. After publication [1,2] appears the greater amount of work, in which were investigated the properties of materials with negative index of refraction and were done the attempts to evaluate their possible practical applications. Very good selection of texts of these work possible to find on the site of University San-Diego [4].

 The Materials, which possess negative index of refraction, there were is named in [1,2] "Left-handed materials". This term, well sounding on english, has not the euphonious translation on russian, and so we shall name the materials with **n<0** "materials with negative refraction",  for shprt squall "MNR" ("ВОП" in russian). Accordingly, usual materials with **n>0** possible to mark as "MPR" ("ВПП" in russian). The term "MNR" squall in some measure corresponds to the term "negative refraction", which also often is used in english publications on givenned themes.

The appearance MNR, generally speaking, does not bring about appearance some absolutely unusual phenomenas, but will easy make sure, that in the case of MNR some optical laws are realized otherwise than in accustomed for us MPR. These differences for Snellius law, Doppler effect and Cherenkov  effect there were are considered in [3]. In that place, there was is shown that proturberant and concave lenses as it were "are changed the places", but usual flat slab can under some conditions to be collecting lens, as this shown on fig.1. It is enough full consideration such flat lens is kept in [5].

To the of discussed in [3] phenomenas and laws necessary to add one more important law, more exactly principle - Fermat`s principle. The wording of this principle meets in literature in two different variants[1]:

---

[1] The wording of Fermat`s principle in "Physical encyclopedic dictionary", M., Soviet Encyclopedia, 1983. The article "Fermat`s principle": most simplest form F.p., - statement that ray of light always spreads in space between two points on that way, along wich time of its passing less, than along any one of others ways, connecting these points.
Wording of Fermat`s principle in British encyclopedia, article "Fermat`s principle" (http://www.britannica.com/search?query=Fermat%60s%20principle&ct=&fuzzy=N): statement that light traveling between two points seeks a path such that the number  of waves (the optical length between the points) is equal, in the first approximation, to that in neighbouring paths. Another way of stating this principle is that the path taken by a ray of light in traveling between two points requires either a minimum or a maximum time.

1. The light spreads from one point of space in another on most short path. Here, term " most short path " implies the minimum expenses of time for passing of this way.
2. The light spreads from one point of space in another on paths, which corresponds to the minimum length of optical way. Under term "optical way" is understood the distance, which light pass in vacuum for time of spreading from one point of space to another.

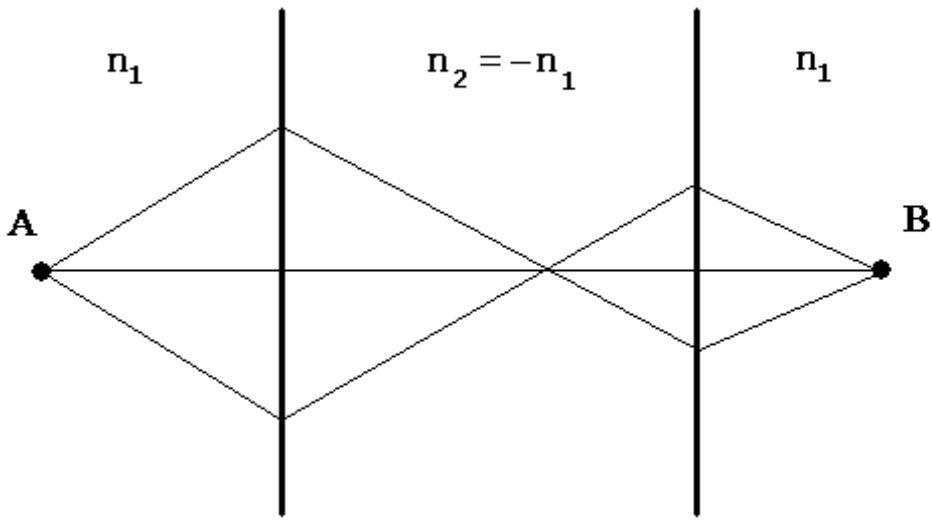

Fig.1. Passing of light from object A to image B through flat lens, made from MNR with index of refractions $n_2 = - n_1$

Except these differences, often becomes firmly established (absolutely correct) that term "minimum of way or time" in some cases should be replaced on term "maximum" or even simply "extremum".

If now return to our two afore-cited wordings, that immediately becomes clear that they both equally correct for the case, when light passes through MPR only, but they both are not correct, when way of spreading the light lies partly through MPR, partly through MNR. This is easy confirmed the drawing 2, on which are shown possible ways of ray, crossing flat surface, separating two media, having index of refraction $n_1$ and $n_2$, respectively.

In that event if $n_1$ and $n_2$ are both positive (that is to say both half-space consist of MPR), ray goes on way $AO_1B$, and angles φ and ψ satisfy Snellius law

**sinφ $n_1$ = sinψ $n_2$**       (1)

Optical length of this way **L** is equal:

$$L = n_1 \cdot (AO_1) + n_2 \cdot (O_1 B) \tag{2}$$

Not difficult show that Snellius law (1) will be valid in that, and only in that case if variation of optical lenght (2) $\delta L$ is zero

$$\delta L = \delta \{n_1 \cdot (AO_1) + n_2 \cdot (O_1 B)\} = 0 \tag{3}$$

Herewith value L itself for real way AO1B will is minimum and positive.

In that event if both values $n_1$ and $n_2$ are negative (i.e. and overhand and from below from borders of section are MPR), move of rays will such, as in previous event, with one important difference. In first event wave vector in both media is directed along rays that is to say from **A** to **B**, but in the second event wave vector is directed against direction of rays that is to say from **B** to **A** [3].

Herewith, optical length L turns out to be negative, and for real way **AO$_1$B** it will is maximum.

Both considered event correspond to positive values **n=n2/n1>0** that is to say relative factor of refraction of second media comparatively the first.

The position greatly changes if value n=n2/n1 turns out to be negative. This

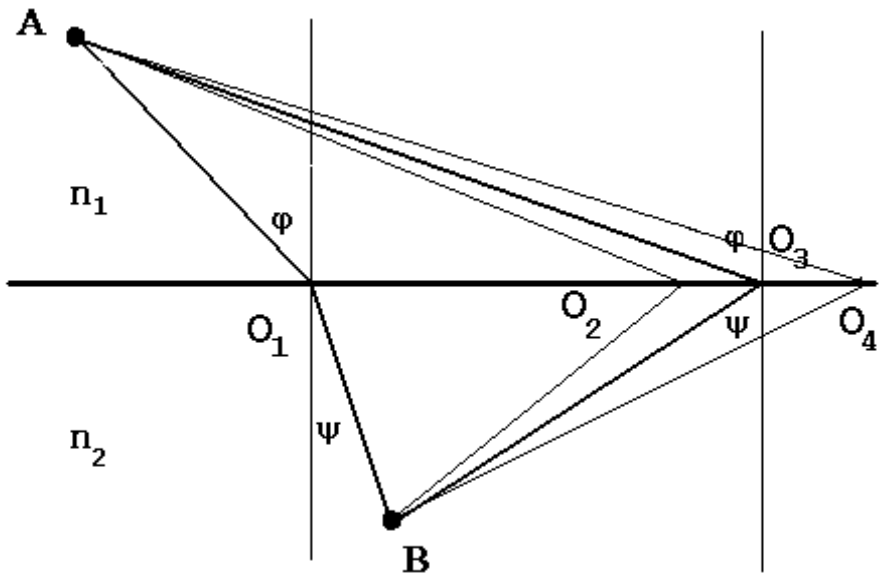

Рис.2. The passing of light from point A to point B through flat border between two media with refraction indexes $n_1$ and $n_2$, respectively.
- Case **n=n$_2$/n$_1$>0**. The light spreads on way **AO$_1$B**.
- Case **n=n$_2$/n$_1$<0**. The light spreads on way **AO$_3$B**. The ways **AO$_2$B** and **AO$_4$B** - are virtual ways of spreading the light for this case.

will be in that event if on one side borders of section inheres MPR, but on other - MNR. In this case ray from first ambience in the second will go on way **AO₃B**, and for angles j and y will be executed the Snellius law (1), but already under negative value of angle y. For real way of spreading will fair be correlation (4)

**δL=δ{n₁*(AO₃)+ n₂*(O₃B)} =0**                          (4)

This correlation changes (3) at substitution in it negative value **n** for MNR, for instance **n2<0**. Hereunder for real way is executed the condition "extremum its optical length" but this length is defined through index of refraction taking into account its sign. However in this case it is impossible a priori confirm that real way of light corresponds to exactly to maximum or exactly minimum of optical way. The type of extremum in this instance depends on geometries of problem and concrete values **n₁** and **n₂**.

Much important that in this case real way from point **A** to point **B** is not shortest on time of spreading. So, virtual way **AO₂B** will is passed the light for smaller time, but way **AO₄B** - for greater time than time of passing by light of real way **AO₃B**.

Thereby wording of Fermat`s principle through time of spreading the light in general event is not correct. The correct wording is only wording of this principle through extremum of length of optical way:

The real way of spreading the light in ambience corresponds to the local extremum of length of optical way.

Using the term "local" takes into account the fact that in problem can be several possible optical ways, such that for they are executed condition (3) and (4).

Length of optical way **L** between points **A** and **B** in most commonly case, when index of refractions **n** is changed from point to point, is equal to integral

$$L = \int_A^B n\,dl \qquad (5)$$

Since value **n**, falling into (5) can be negative, that clear that length of optical way **L** (really this value is an eiconal) can have any sign and any value. So, this length will be negative if light passes through MNR only. In some cases length of optical way can be equal the zero. Exactly such a zero length of optical way between object and its image is in lens made from NMR, showed on fig.1 [5]. The concept of optical lenght is connected with total phase wind of wave, which depends on the index of refraction **n**, which defines the phase velocity of light, rather then group. Often using determination of length of optical way as a time of spreading the light on essences of deal identifies phase and group velocity, that untrue in general case, and for MNR gives very rough mistakes.

The difference between phase and group velocity for event of lens, showed on fig.1, brings else to one effect. This is because time of spreading the light on central ray and on peripheral rays in this device turns out to be different though optical length for all rays alike. So at passing through such lens of ultrashort pulses of light their form will be distorted. This defect do not possess (in ideal) usual lenses, made from MPR.

The work has been funded by Russian Foundation of Basic Research in accordance with grant # 01-02-16596a